\documentclass[12pt]{iopart}
\usepackage{epsfig,amssymb,subfigure,float}
\begin{document} 
\newcommand {\cs}{$\clubsuit$}
%\renewcommand{\baselinestretch}{2} 
%\begin{frontmatter}
\title{Boson-fermion mixtures inside an elongated cigar-shaped trap}
\author{Z. Akdeniz$^{1,2}$, P. Vignolo$^1$\footnote{Author to whom any correspondence should be addressed ({\tt vignolo@sns.it})} and M. P. Tosi$^1$}
\address{$^1$NEST-INFM and Classe di Scienze, Scuola Normale Superiore,
I-56126, Pisa, Italy\\
$^2$Department of Physics, Istanbul University, Istanbul, Turkey}

\begin{abstract}
We present mean-field calculations of the equilibrium state in a 
gaseous mixture of bosonic and spin-polarized fermionic atoms
with repulsive or attractive interspecies interactions, confined 
inside a cigar-shaped trap under conditions such that the radial thickness of 
the two atomic clouds is approaching the magnitude of the 
$s$-wave scattering lengths. 
In this regime the kinetic pressure of the fermionic component
is dominant. Full demixing under repulsive boson-fermion
interactions can occur only when the number of fermions in the
trap is below a threshold, and collapse under attractive interactions 
is suppressed
within the range of validity of the mean-field model.
Specific numerical illustrations are given for values of system parameters
obtaining in $^7$Li-$^6$Li clouds. 
\end{abstract}

\vspace{0.5cm}
%\begin{keyword} 
%Bose-Fermi gases \sep quantum phase separation \sep equilibrium properties  
%\end{keyword}
\maketitle
%\end{frontmatter}

\section{Introduction}
Dilute boson-fermion mixtures have been studied
in several experiments by trapping and cooling gases of mixed
alkali-atom isotopes 
\cite{Schreck2001a,Goldwin2002a, Hadzibabic2002a,Roati2002a,Modugno2002a,Ferlaino2003a}.
The boson-fermion coupling strongly affects the equilibrium properties
of the mixture and can lead to quantum phase
transitions (for a review see \cite{bf_review}). In particular,  
boson-fermion repulsions in three-dimensional (3D) clouds can induce 
spatial demixing of
the two components when the interaction energy overcomes the kinetic
and confinement energies \cite{Molmer1998a}. 
This occurs in 3D if the fermion density is high 
or if the boson-fermion repulsion is strong, as shown in a
phase diagram obtained 
by imposing pressure balance and equality of chemical potentials
in coexisting phases \cite{Viverit2000a,Akdeniz2002a}. 
Although spatial demixing has not yet
been experimentally observed, the experiments of Schreck
{\it et al.}  \cite{Schreck2001a} on $^7$Li-$^6$Li mixtures inside
an elongated 3D trap appear to be close to the
onset of a demixed state \cite{Akdeniz2002b}.
Collapse driven by increasing the number of bosons in a 3D trap has instead
been observed in $^{87}$Rb-$^{40}$K mixtures \cite{Modugno2002a},
where the interspecies interactions are attractive.

On approaching a quasi-two-dimensional (Q2D) situation, 
the Fermi energy starts to increase linearly with the number 
of fermions per unit area, so that the linear stability condition
of the mixture becomes 
independent of the fermion density and involves only the scattering 
lengths and the transverse width of the cloud.
In this regime spatial demixing of a $^7$Li-$^6$Li gas is predicted
to occur by simply squeezing down the axial thickness of the atomic 
clouds \cite{Akdeniz2004a}.

The analysis of Viverit \textit{et 
al.} \cite{Viverit2000a} for homogeneos 
boson-fermion matter has been extended by 
Das \cite{Das2003a} 
to a one-dimensional (1D) homogeneous mixture. 
In the 1D case the Fermi energy is proportional to the square 
of the number of fermions per unit length, so that the mixed 
state becomes unstable at {\it low} linear fermion densities. 
The dependence of the kinetic pressure of the Fermi gas
on the fermion density 
becomes stronger with decreasing dimensionality, and in 1D a 
number of fermions above a threshold will want to occupy the 
whole available space irrespectively of the repulsions exerted 
by the bosons. Similarly, it is easily shown that
a 1D homogeneous mixture with
attractive boson-fermion 
interactions is stabilized against collapse by the Fermi pressure 
if the linear density of fermions is above the same threshold.

In this paper we examine the instabilities
of a dilute boson-fermion mixture trapped 
inside a strongly elongated harmonic confinement. 
We give the general conditions under which 
various forms of demixing can occur in
a Q1D confined geometry  in terms of the physical parameters of the system
and draw a phase diagram at zero temperature in a plane defined by
two scaling parameters.
We support our analysis by extensive mean-field calculations
of the equilibrium density profiles in a $^7$Li-$^6$Li
mixture. We find
that in the case of interspecies repulsions the stability
condition which applies in the macroscopic limit
is still approximately valid for full demixing in harmonic confinement.
We next turn to
the case of attractive interspecies interactions,
where we search for a collapse instability by a 
variational approach patterned after the work carried out by
Miyakawa {\it et al.} \cite{Miyakawa2001a}
on a 3D mixture.
We find that our mean-field model does not allow collapse in a 
confined Q1D geometry, but only predicts that the density profiles keep 
narrowing with increasing attractive coupling strength until the validity
of the model is lost. These predictions of the variational analysis are 
again supported by numerical illustrations of density profiles.

The paper is organized as follows. In Sec. \ref{method} we describe the
mean-field model and report the stability condition
for a macroscopic mixture in 1D and Q1D geometry. 
Section \ref{secdem} discusses the
conditions under which various forms of demixing
occur in a Q1D harmonic confinement and shows some illustrative
configurations of the density profiles in the gaseous cloud
in the phase-separated regime.
The variational analysis
and some illustrative density profiles for 
the case of interspecies attractions
are given in Sec. \ref{seccoll}.
Finally, Sec. \ref{secconcl} gives a summary and some concluding remarks. 

\section{The model and the phase diagram} \label{method}
The atomic clouds are trapped in cigar-shaped potentials given by
\begin{equation}
\hspace{-2cm}V^{ext}_{b,f}(x,y,z)=m_{b,f}
\omega^2_{\perp(b,f)}(x^2+ y^2)/2+m_{b,f}\omega^2_{z(b,f)}z^2/2
\equiv U_{b,f}({x,y})+V_{b,f}(z)
\end{equation} 
where $m_{b,f}$  are the atomic masses and
$\omega_{z(b,f)}\ll\omega_{\perp(b,f)}$ the trap frequencies.
We assume repulsive interactions between the bosons and
focus on the case where the trap is elongated enough that 
the dimensions of the atomic clouds in the radial plane, which are of the 
order of the
radial harmonic-oscillator lengths 
$a_{\perp(b,f)}=(\hbar/m_{b,f}\omega_{\perp(b,f)})^{1/2}$, are comparable
to the magnitude of the
3D boson-boson and boson-fermion scattering lengths, $a_{bb}$ and $|a_{bf}|$.
In this regime we can study the equilibrium properties of the mixture
at essentially zero temperature ($T\simeq 0.02\,T_F$)
in terms of the particle density 
profiles along the $z$ axis, 
which are $n_{c}(z)$ for the Bose-Einstein condensate
and $n_{f}(z)$ for the
fermions. The profiles are evaluated
using the Thomas-Fermi approximation for the condensate and the
Hartree-Fock approximation for the spin-polarized fermions.
Here and in the following 
we shall take $a_{\perp b}=a_{\perp f}$ $(=a_\perp$, say).

The Thomas-Fermi approximation assumes that the number of condensed
bosons is large enough that the kinetic energy term in the
Gross-Pitaevskii equation may be neglected~\cite{Baym1996a}. It yields  
\begin{equation}
n_c(z)=[\mu_b-V_b(z)-g_{bf}n_f(z)]/g_{bb}
\label{zehra1}
\end{equation}
for positive values of the function in the square brackets, and zero
otherwise. Here, $\mu_b$ is the chemical potential of the bosons.
This mean-field model is valid when the high-diluteness 
condition $n_ca_{bb}\ll 1$ holds and the temperature is outside the critical 
region.
If the conditions $a_\perp>  (a_{bb},|a_{bf}|)$
are fulfilled, the atoms experience collisions in three dimensions
and the coupling constants in Eq. (\ref{zehra1}) can be written in terms 
of those for a 3D cloud as~\cite{salasnich,Akdeniz2003a}
\begin{equation}
g_{bb}=\frac{g_{bb}^{3D}}
{2\pi a_\perp^2}\;,\;\;\;g_{bf}=\frac{g_{bf}^{3D}}{2\pi a_\perp^2}  
\label{1Dint}
\end{equation} 
with $g_{bb}^{3D}=4\pi\hbar^2a_{bb}/ m_b$, 
$g_{bf}^{3D}=2\pi\hbar^2a_{bf}/m_r$ and
$m_r=m_bm_f/(m_b+m_f)$.

The Hartree-Fock
approximation~\cite{Minguzzi1997a,Amoruso1998a,Vignolo2000b}
treats the fermion cloud as an ideal gas subject to an effective 
external potential, that is 
\begin{equation}
n_{f}(z)=\int\frac{dp}{2\pi\hbar}\left\{\exp\left[\left(
\frac{p^2}{2m_{f}}
+V^{eff}(z)-\mu_{f}\right)/k_BT\right]+ 1\right\}^{-1},
\label{zehra4}
\end{equation}
where $\mu_f$ is the chemical potential of the fermions and
\begin{equation}
V^{eff}(z)=V_{f}(z)+g_{bf}n_c(z).
\label{zehra3}
\end{equation}
The fermionic
component has been taken as a dilute spin-polarized gas, for which the
fermion-fermion  $s$-wave scattering processes are inhibited  by
the Pauli principle and $p$-wave scattering is
negligibly small~\cite{Demarco1999b}. 
In the mixed regime the 
Fermi wave number in the axial direction and
the condensate density should be smaller than $1/|a_{bf}|$, 
but this is not
a constraint in the regime of demixing where the boson-fermion
overlap is rapidly dropping. 
On the contrary, for large attractive interactions
this condition, together with the diluteness condition
$n_c a_{bb}\ll1$, may not be fulfilled and this sets the range of validity
of our results.

The chemical potentials $\mu_{b,f}$  characterise the system in the 
grand-canonical ensemble and are determined by requiring that
the integrals of the densities over the $z$ axis 
should be equal to the  average numbers $N_b$ and $N_f$ of particles.
The presence of a bosonic thermal cloud can be 
treated by a similar Hartree-Fock approximation \cite{Akdeniz2002b}, 
but it has quite negligible effects at the 
temperatures of present interest.

As the boson-fermion coupling increases, the 
mixture can become unstable against demixing 
(in the case $g_{bf}>0$) or collapse (in the case $g_{bf}<0$).
In the macroscopic 1D limit at given particle densities a linear
stability analysis based on a mean-field energy functional predicts 
that the locations for
demixing and collapse coincide and that the gaseous cloud is stable
if the usual condition
\begin{equation}
g_{bb}g_{ff}-g_{bf}^2\ge0
\label{condhom}
\end{equation}
is fulfilled \cite{Viverit2000a},
with $g_{ff}=\hbar^2\pi^2n_f/m_f$ playing the role of an 
effective fermion-fermion repulsion
due to the Pauli pressure in the Fermi gas.
Since $g_{ff}$ increases with $n_f$, the mixed state
is stable at given boson-fermion attractive or repulsive coupling strength
if the linear density of fermions is {\it above} a threshold \cite{Das2003a}.

\begin{figure}
\centerline{
\epsfig{file=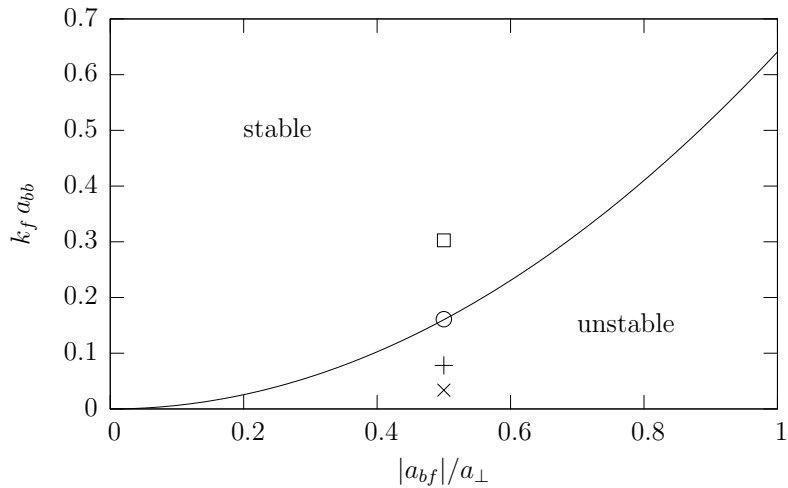,width=0.7\linewidth}}
\vspace{0.2cm}
\caption{Phase diagram of a Q1D $^7$Li-$^6$Li mixture
at $T=0$, in the plane defined
by the dimensionless parameters $a_{bb}k_f$ and $|a_{bf}|/a_\perp$.
The symbols at $|a_{bf}|/a_{\perp}=0.5$ correspond to the parameters of the
configurations shown in Figs. \ref{fig_dem}
and \ref{fig_coll}: ($\square$) $N_f=3\times 10^5$, (O) 
$N_f=8,5\times 10^4$, (+) $N_f=2\times 10^4$, ($\times$) $N_f=3740$.}
\label{fig3}
\end{figure}

The stability condition in Eq. (\ref{condhom}) can be extended to the Q1D
geometry for large numbers of particles and if the boson and fermion densities
vary smoothly (for the sake of brevity in the following we will refer 
to this regime as the ``Q1D macroscopic limit''). 
By using Eq.(\ref{1Dint}) 
for the boson-boson and boson-fermion couplings and by taking the 
value of $n_f$ as the linear fermion density at the centre of the trap,
Eq. (\ref{condhom}) is rewritten as
\begin{equation}
\frac{|a_{bf}|}{a_\perp}\le \left(\frac{2\pi m_r^2}{m_b m_f}\,k_f a_{bb}
\right)^{1/2}
\label{condition}
\end{equation}
where $k_f=(2N_f)^{1/2}/a_{zf}$ is the Fermi wave number 
under harmonic confinement at the centre of the trap, with $a_{zf}$ the axial
oscillator length for the fermions.
In Fig. \ref{fig3} we have plotted Eq. (\ref{condition}) 
in the plane 
$\{|a_{bf}|/a_\perp,\, k_f a_{bb}\}$ for the case of a $^7$Li-$^6$Li mixture.

The effects of finite size and dishomogeneity of the
gaseous clouds are not taken into account in Eq. (\ref{condition}),
with the consequence that the resulting phase diagram
in Fig. \ref{fig3} shows a sharp transition as in
the  macroscopic limit.
Transitions in a mesoscopic cloud under confinement are instead 
spread out and several alternative definitions of their location 
can therefore be given. These are discussed in the next two sections.
We shall for the sake of brevity continue to use expressions 
such as ``phase transitions'' and ``phase diagram'' when needed.

\section{Spatial demixing}
\label{secdem}
In the calculations that we report in the following
we have used 
values of the trap frequencies
appropriate to the experiments in Paris on the
$^7$Li-$^6$Li mixture \cite{Schreck2001a}, that is 
$\omega_{zb}/2\pi=4000$~Hz and $\omega_{zf}/2\pi=3520$~Hz.
We also take $a_{bb}=5.1\,a_0$, $a_0$ being the Bohr radius, 
as appropriate to $^7$Li-$^7$Li
$s$-wave scattering.
We first discuss in this Section 
three alternative locations of the demixing point in
a mesoscopic trapped cloud, that we denote as partial, dynamical, 
and full demixing. Their definition and the derivation of simple 
analytical expressions in the Q1D model are given below.

\subsection{Partial demixing}
\label{partial_sep}
\begin{figure}[H]
\centerline{
{\epsfig{file=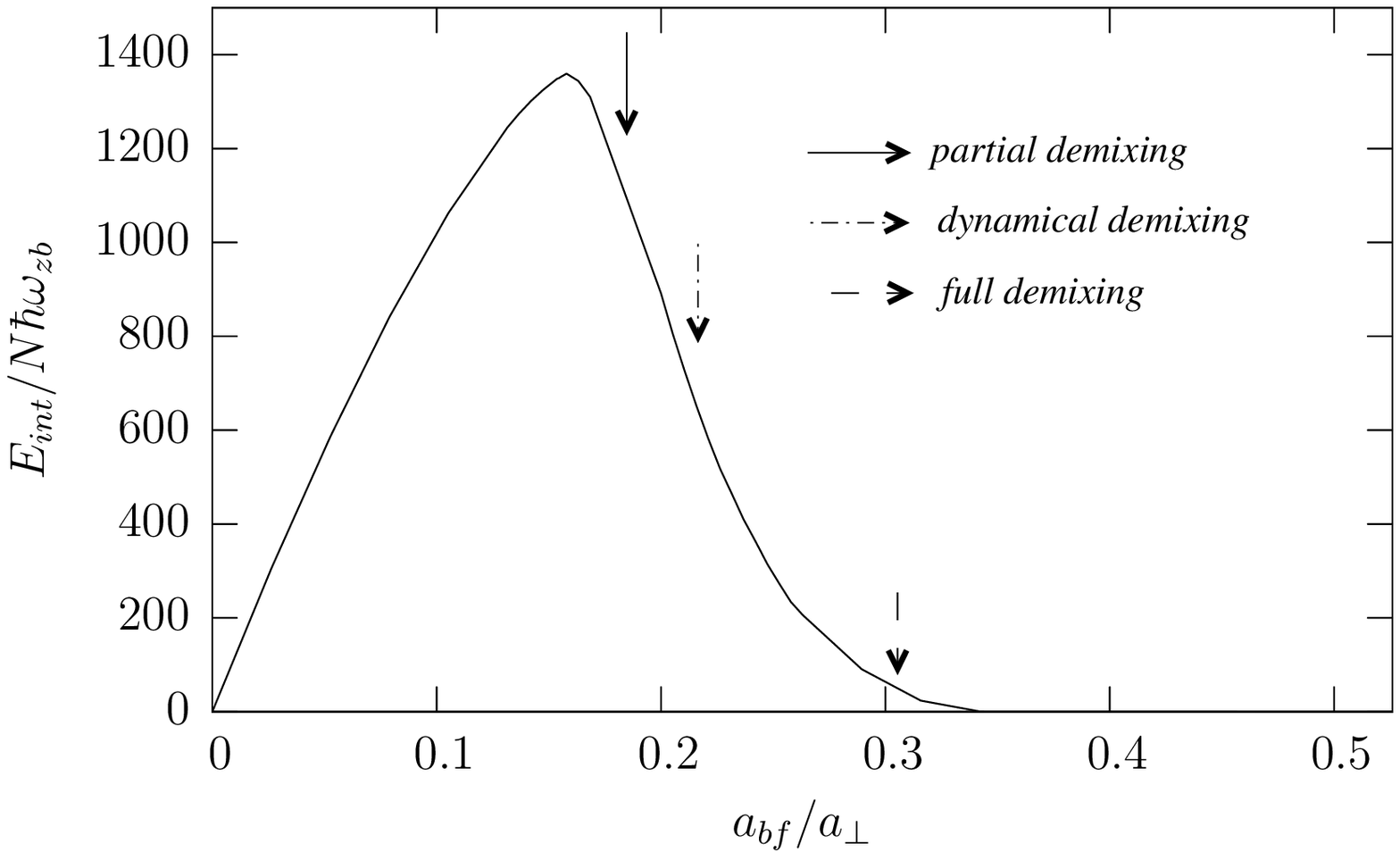,width=0.48\linewidth}}
{\epsfig{file=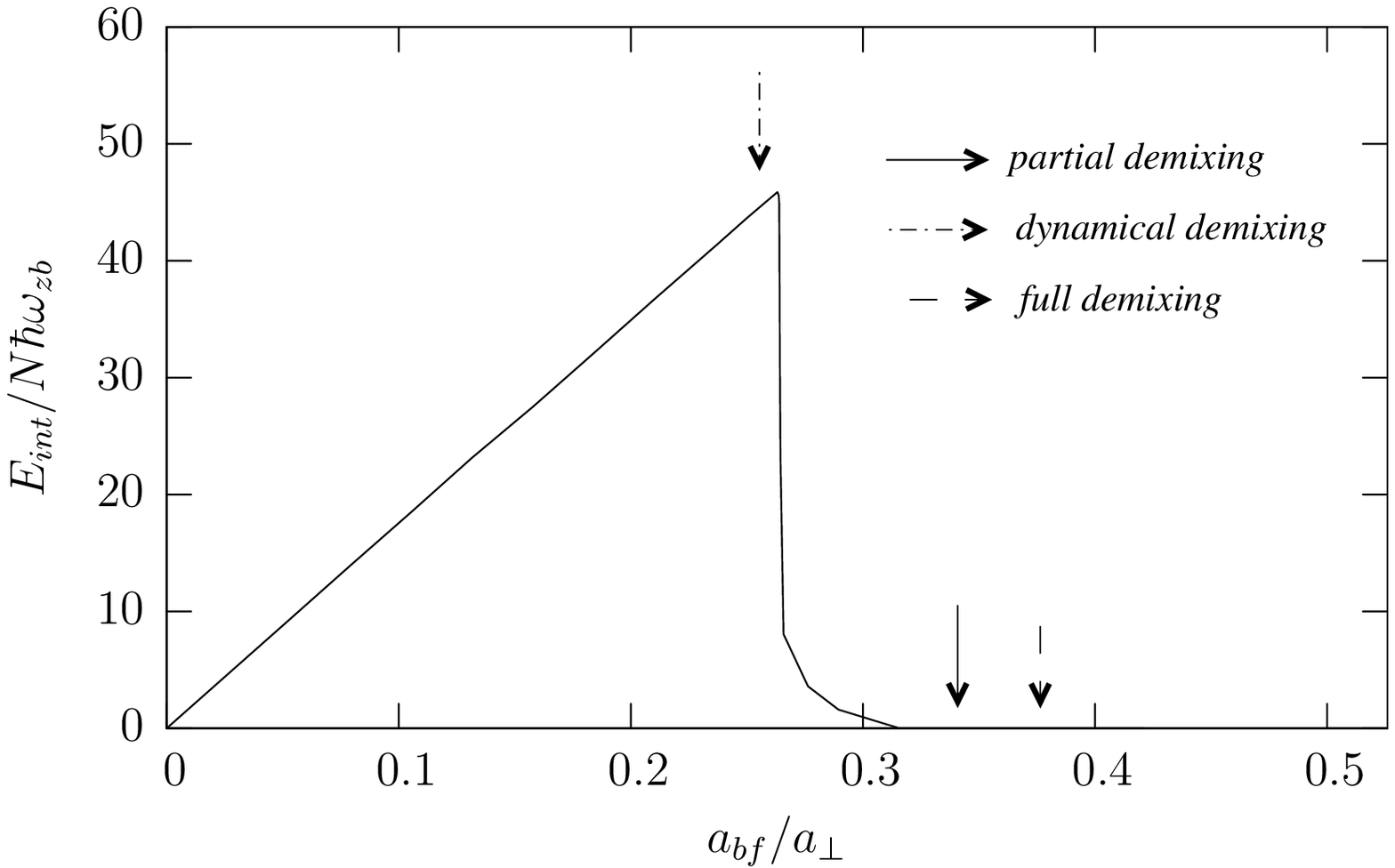,width=0.48\linewidth}}}
\caption{Boson-fermion interaction energy $E_{int}$ 
(in units of $N\hbar\omega_{zb}$, with $N=N_b+N_f$), 
as a function of $a_{bf}$ (in units of $a_\perp$) for $a_{bb}=5.1\,a_0$.
Left panel: $N_b=N_f=10^4$. Right panel: $N_b=100$ and $N_f=19900$. 
The arrows indicate the locations of demixing
estimated from Eqs. (\ref{partially0}), (\ref{dyn0}), 
and (\ref{condition_full}).}
\label{fig2}
\end{figure}

The interaction energy $E_{int}$ between the boson and fermion clouds 
initially grows as the boson-fermion coupling is increased, 
but in the demixing regime reaches a maximum and then 
falls off as the overlap between 
the two separating clouds diminishes. 
We locate partial demixing at the maximum of 
$E_{int}$, that we calculate from the density profiles according to 
the expression
\begin{equation}
E_{int}=g_{bf} \int dz \, n_c(z) n_f(z).
\end{equation}
The behaviour of the interaction energy as a function of 
$a_{bf}$ at
$a_{\perp}=38\,a_0$, and for various numbers of bosons and fermions, 
is shown in Fig. \ref{fig2}.
An approximate analytic formula for the location of partial demixing 
as a function of the system parameters in the Q1D model can be 
obtained from the condition 
$\partial E_{int}/\partial g_{bf}=0$ which leads to the relation
$n_c(z)n_f(z)-n_f^2(z)/g_{bb}-n_c^2(z)/g_{ff}=0$. By 
using the estimate $n_{c,f}\approx N_{b,f}/(2 R_{b,f})$ 
with the values of the cloud radii in the absence of 
boson-fermion interactions,
$R_f=(2 N_f)^{1/2}a_{zf}$ and 
$R_b=[3N_b g_{bb}a_{zb}/(2\hbar\omega_{zb})]^{1/3}a_{zb}$, 
where $a_{z(b,f)}=(\hbar/m_{b,f}\omega_{z(b,f)})^{1/2}$,
the location of the maximum in the boson-fermion interaction 
energy as a function of $a_{bf}/a_\perp$ is found to lie at
\begin{equation}
\left.\frac{a_{bf}}{a_\perp}\right|_{part}\simeq\left(c_1\, k_f a_{bb}+
\frac{c_2}{k_f^2 a_{bb}^2}\right)^{-1}
\label{partially0}
\end{equation}
where we have defined
\begin{equation}
c_1=\frac{1}{4N_b^{2/3}}\frac{m_b a_{zb}}{m_r a_\perp}
\left(\frac{3a_{bb}a_{zb}}
{a_\perp^2}\right)^{1/3},
\end{equation}
\begin{equation}
c_2=\frac{8N_b^{2/3}}{\pi^2}\frac{m_f a_{bb}}{m_r a_{zb}}
\left(\frac{a_{bb}^2}{3a_\perp a_{zb}}\right)^{1/3}.
\end{equation}
The prediction obtained from Eq.~(\ref{partially0}) is shown in
Fig. \ref{fig2} by solid arrows.
For the case $N_b=N_f=10^4$, there clearly is reasonable
agreement  between the analytical estimate and the numerical results.
For a low number of bosons and a large number of fermions the condition
given in Eq. (\ref{partially0}) is dominated by the term involving $c_1$.
In this limit, the bosonic density is largely deformed by the pressure of
 the fermions and the
prediction obtained from Eq.~(\ref{partially0}) overestimates the
position of the maximum of the interaction energy
(see right panel in Fig. \ref{fig2}).

\subsection{Dynamical demixing}
At some value of the boson-fermion coupling 
the fermion density vanishes at the centre of the trap. This occurs when
\begin{equation}
\left.\frac{g_{bf}}{g_{bb}}\right|_{dyn}=\frac{\mu_f}{\mu_b}.
\label{pablo}
\end{equation}
We denote this point as the dynamical location of demixing in a 
mesoscopic cloud, since we expect a sharp upturn of the low-lying 
fermion-like collective mode frequencies to occur at this point, where the topology of the cloud changes.
This has been shown to the case for both collisional and collisionless excitations 
in a mixture under 3D confinement ~\cite{Capuzzi2003a,Capuzzi2003b}.

If we insert the chemical potentials for ideal-gas clouds in 
Eq. (\ref{pablo}), we obtain an 
approximate expression for the location of dynamical demixing,
\begin{equation}
\left.\frac{a_{bf}}{a_\perp}\right|_{dyn}\simeq
\left(\frac{\pi m_r^2}{m_b m_f}\,k_f a_{bb}
\right)^{1/2}.
\label{dyn0}
\end{equation}
The prediction obtained from Eq.~(\ref{dyn0}) is indicated in
Fig.~\ref{fig2} by dot-dashed arrows.
For both small and large values of the number of bosons,
the dynamical condition for demixing is in good agreement with the
numerical results from the boson-fermion interaction energy.

\subsection{Full demixing}
\label{total_PS}
The point of full demixing is reached when the boson-fermion overlap 
becomes negligible as in a phase transition occuring in a
macroscopic cloud. If we use the stability 
criterion given in Eq. (\ref{condition}) for the macroscopic Q1D case, 
the condition for full phase separation at $T = 0$ inside a strongly
elongated trap is given by 
\begin{equation}
\left.\frac{a_{bf}}{a_\perp}\right|_{full}
\simeq \left(\frac{2\pi m_r^2}{m_b m_f}\,k_f a_{bb}
\right)^{1/2}
\label{condition_full}
\end{equation}
or $a_{bf}|_{full}\simeq\sqrt{2}a_{bf}|_{dyn}$.
 The prediction obtained from Eq. (\ref{condition_full}) is indicated 
in Fig. \ref{fig2} by short-dashed arrows.
Analogously to the partial 
demixing condition there is better agreement with the numerical results
on the boson-fermion interaction energy when the number of bosons is 
relative large.

In summary, in Q1D a small number of fermions allows demixing, 
which is in contrast
to the case for full phase separation in 3D
where the critical value of the boson-fermion scattering length
decreases by increasing the number of 
fermions \cite{Viverit2000a,Akdeniz2002a},
and to the Q2D case where the transition is controlled only
by the geometrical parameters of the trap and not by the number
of atoms \cite{Akdeniz2004a}.

The critical coupling strength for both dynamical and full demixing
under harmonic confinement moves upwards with the law
$g_{bf}\propto N_f^{1/4}$ as $N_f$ increases, till
the mixture becomes stabilized by the kinetic pressure of the fermions.

\subsection{Density profiles at full demixing}
\begin{figure}
\centerline{
{\epsfig{file=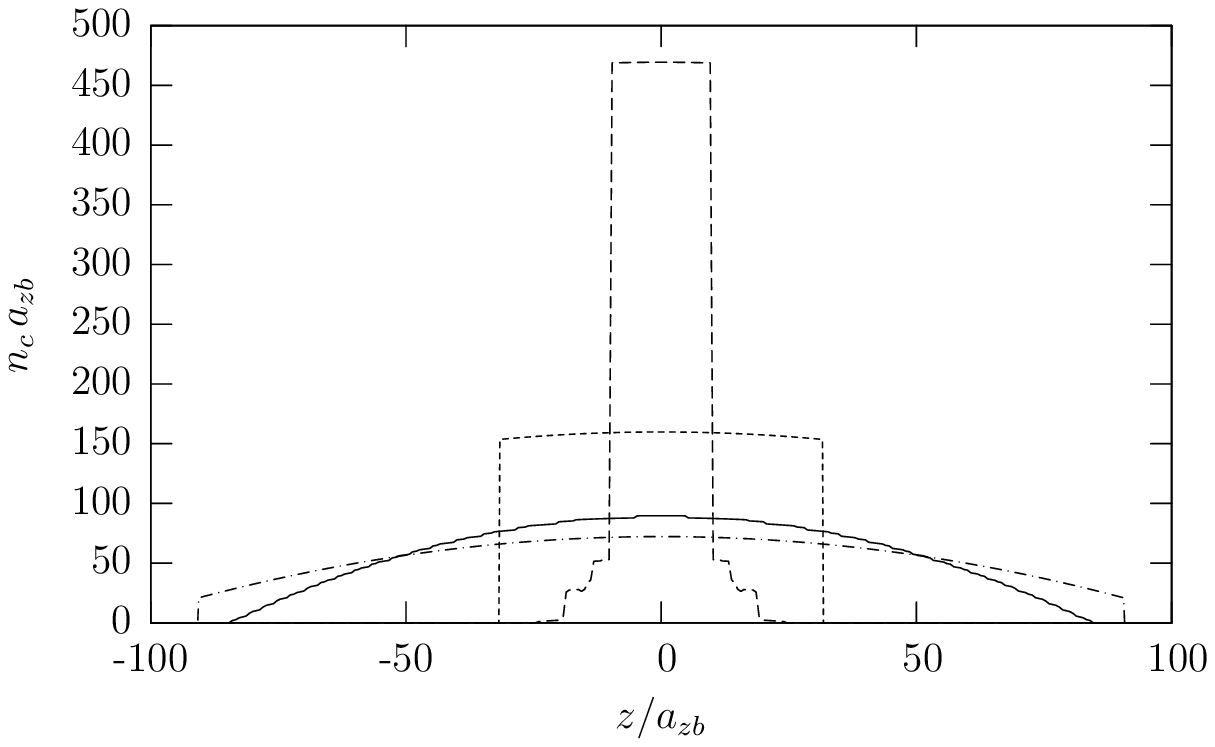,width=0.48\linewidth}}
{\epsfig{file=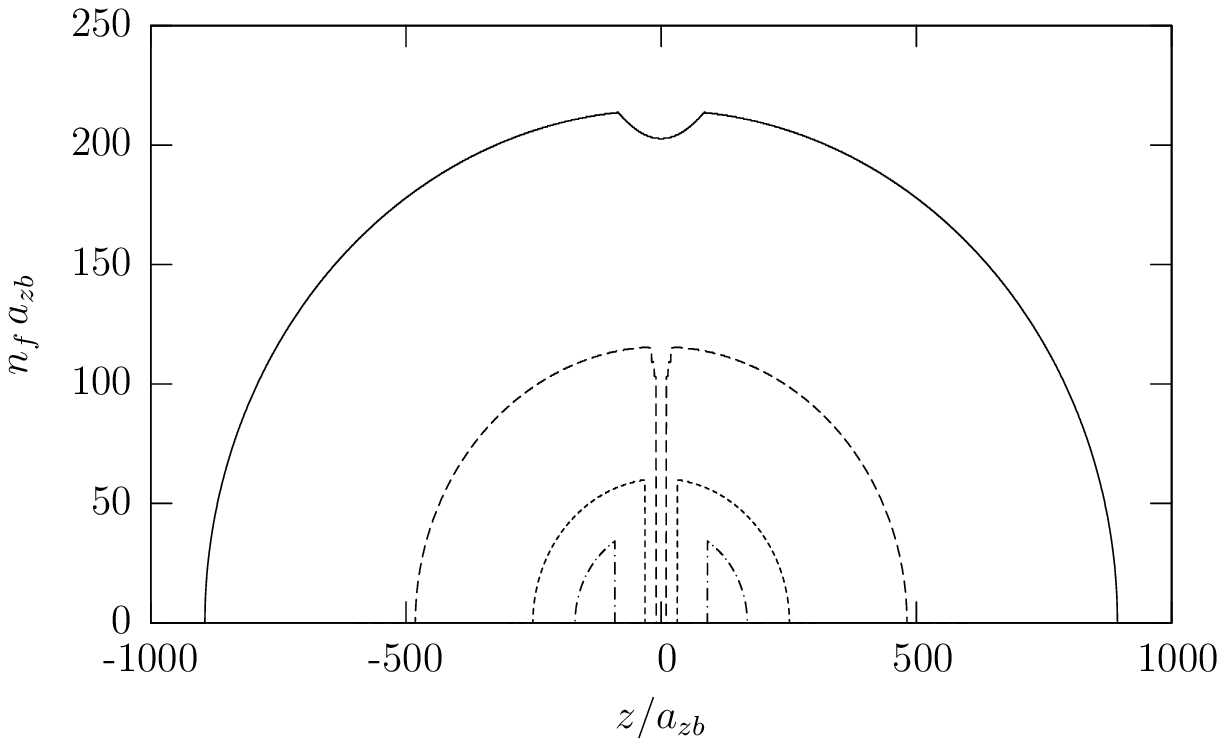,width=0.48\linewidth}}}
\caption{Density profiles of the condensate (left panel) and of the fermions 
(right panel) at
$a_{bf}/a_\perp=0.5$ for various numbers of fermion atoms: 
$N_f=3\times 10^5$ (solid line), $N_f=8.5\times 10^4$ (dashed line), 
$N_f=2\times 10^4$ (short-dashed line), and $N_f=3740$ (dotted-dashed line).
Notice the difference of the horizontal scales in the two panels.} 
\label{fig_dem}
\end{figure}

With the aim of verifying the validity of the condition of full
component separation in Eq. (\ref{condition_full}), 
we have evaluated the boson and fermion density profiles
by varying the boson-fermion scattering length as might be attained 
experimentally by exploiting
optically or magnetically induced Feshbach
resonances \cite{feschbach}, or  by varying $N_f$ {\it i.e.} 
$k_f$ at fixed
radial size $a_\perp=38\,a_0$.
As illustrative examples we show in Fig. \ref{fig_dem}
the density profiles of 
the lowest-energy configurations indicated by the symbols in Fig. \ref{fig3}.
In agreement with the location of the configurations
in the phase diagram, density profiles with a small number of fermions 
are spatially demixed, whereas stability is reached on increasing the 
number of fermions.

\begin{figure}
\centerline{
{\epsfig{file=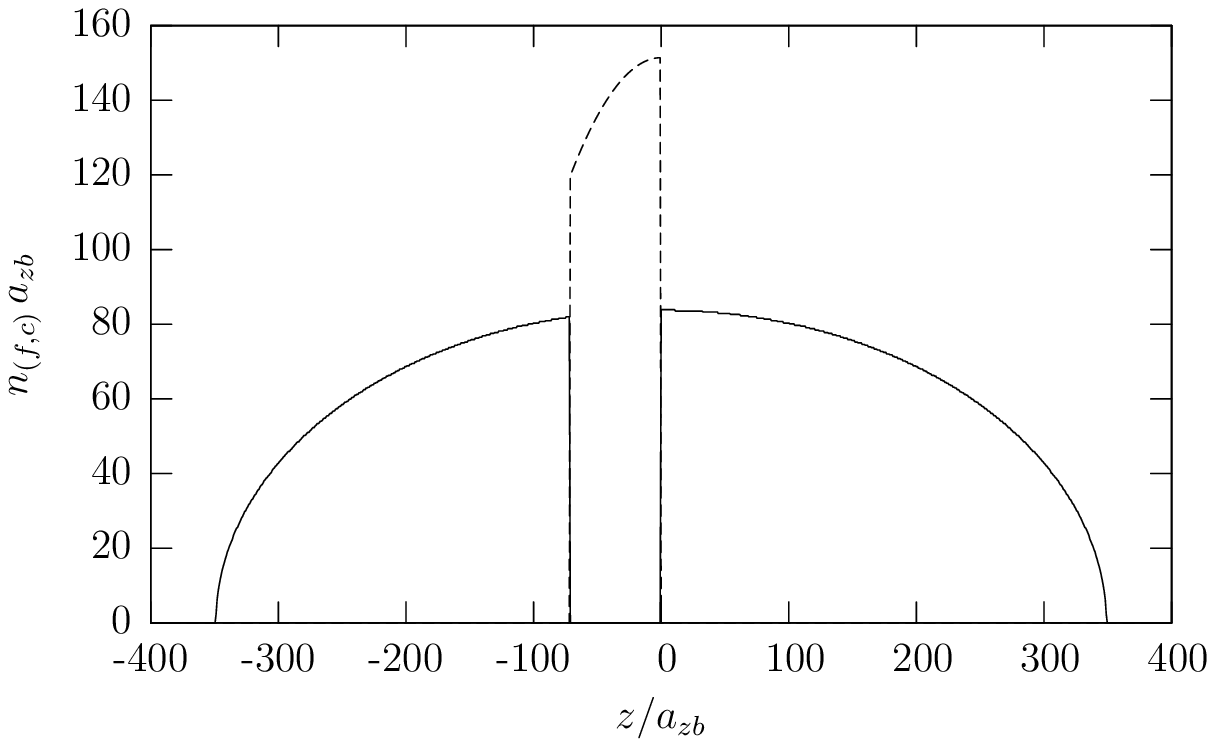,width=0.48\linewidth}}
{\epsfig{file=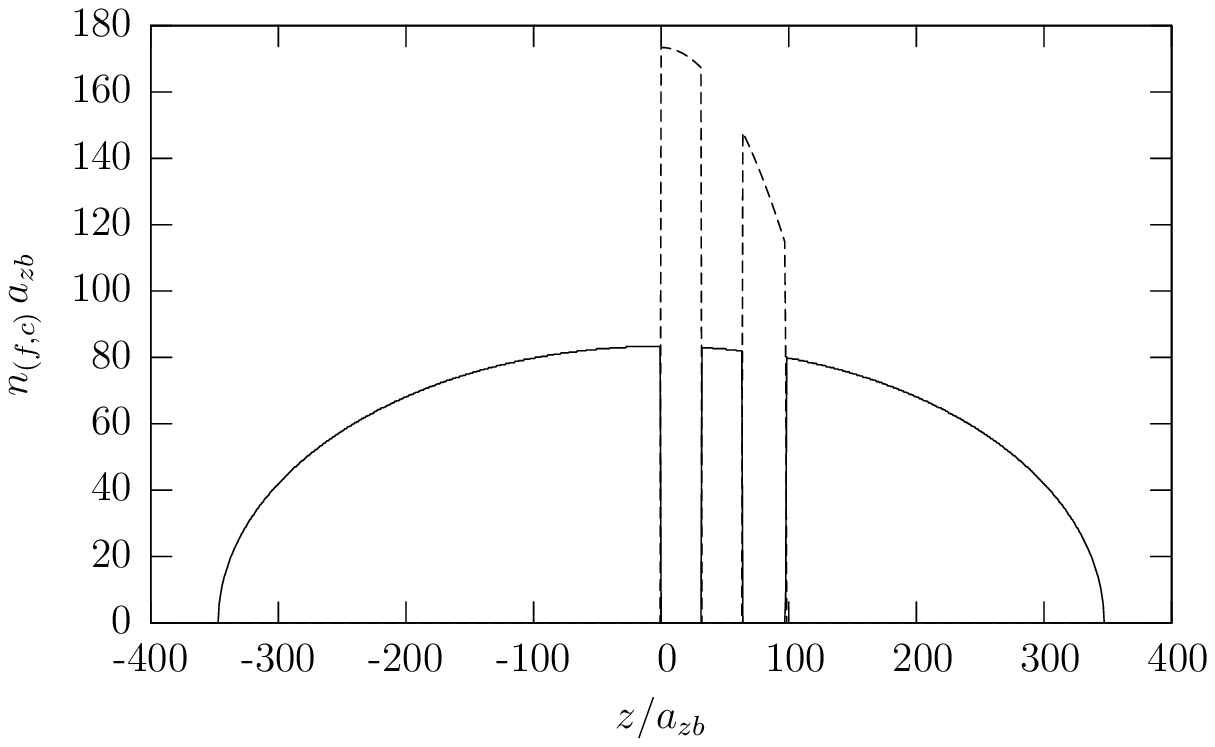,width=0.48\linewidth}}}
\centerline{
{\epsfig{file=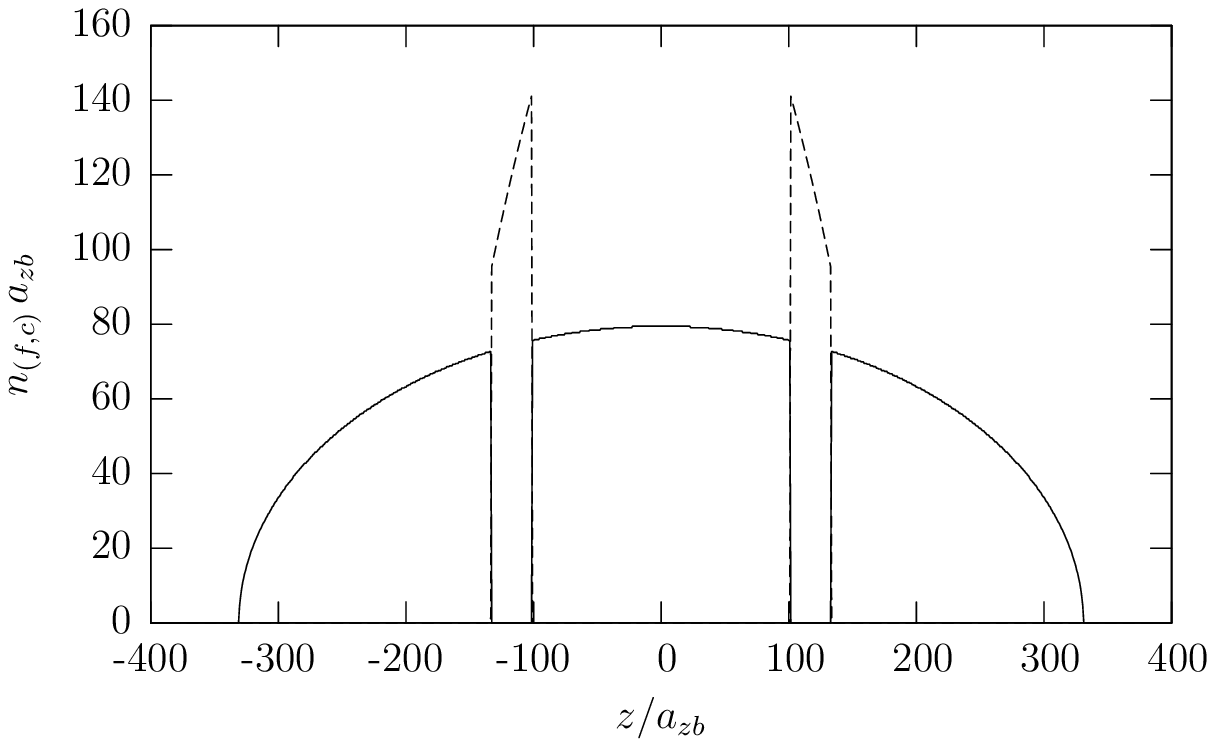,width=0.48\linewidth}}
{\epsfig{file=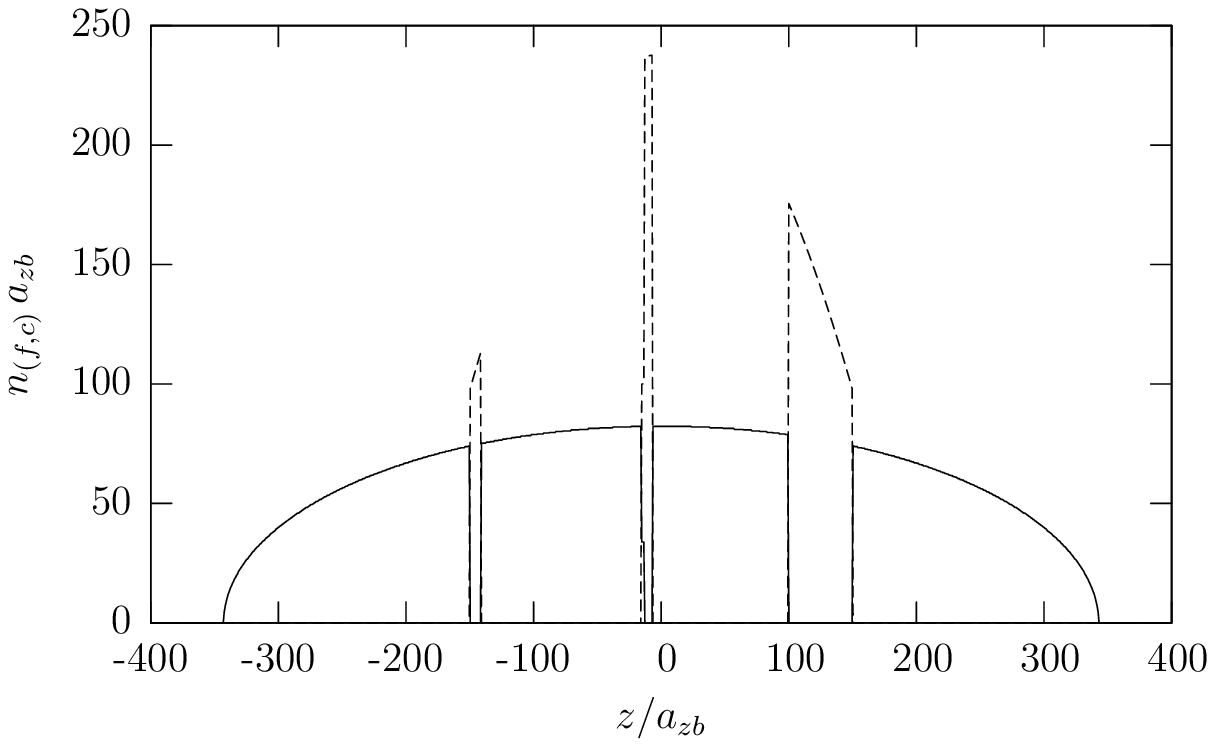,width=0.48\linewidth}}
}
\caption{Metastable 
density profiles of fermions (solid line) 
and bosons (dashed line) at full demixing, 
for a Q1D mixture of $N_b=10^4$ bosons and
$N_f=4\times10^4$ fermions with $a_{bb}=5.1\,a_0$,
$a_{bf}=19 a_0$, and $a_{bf}/a_\perp=0.5$.
The various configurations correspond to 
an excess energy $\Delta E/E$ over the thermodynamically stable demixed
configuration given by
$\Delta E/E\simeq 0.2\%$ 
(left-top panel), $\Delta E/E\simeq 1\%$ 
(right-top panel), $\Delta E/E\simeq 1.6\%$ 
(left-bottom panel), and $\Delta E/E\simeq 5\%$ 
(right-bottom panel).} 
\label{metastable}
\end{figure}

In the region of instability we have also found in addition to the 
thermodynamically stable configurations, shown in Fig. \ref{fig_dem},
various metastable ``exotic''
configurations for the demixed clouds.
Some illustrative examples 
of such metastable density profiles, 
which break axial symmetry and/or consist
of alternating slices of bosons and fermions,
 are
shown in Fig. \ref{metastable}. 
 
\section{Collapse}
\label{seccoll}
In the case of attractive boson-fermion interactions 
we have studied the 
instabilities of the Q1D
harmonically trapped mixture by a variational method.
Following Miyakawa {\it et al.} \cite{Miyakawa2001a}, we write the 
total energy of the gas as a function of the boson order parameter $\Phi(z)$, 
with 
$|\Phi(z)|^2=n_c(z)$, and of the fermion density distribution $n_f(z)$ as
\begin{eqnarray}
&&E[\Phi(z),n_f(z)]=\int dz\left[\frac{\hbar^2}{2m_b}|\nabla\Phi(z)|^2+
V_b(z)|\Phi(z)|^2+\frac{g_{bb}}{2}|\Phi(z)|^4\right]\nonumber\\
&&+\int dz \left[\frac{\hbar^2\pi^2}{6m_f}n_f^3(z)+V_f(z)n_f(z)+g_{bf}n_f(z)|\Phi(z)|^2\right]\,.
\end{eqnarray}
We neglect the kinetic energy of the bosons and make a 
Thomas-Fermi Ansatz for the boson order parameter,
\begin{equation}
\Phi(\tilde z;R)=a_{zb}^{-1/2}\left[\frac{3N_b}{4R}
\left(1-\frac{\tilde z^2}{R^2}\right)\right]^{1/2}
\end{equation}
where $\tilde z=z/a_{zb}$ and the variational parameter $R$ is  
the dimensionless Thomas-Fermi radius of the condensate.
Minimization of the free energy 
$E[\Phi(z),n_f(z)]-\mu_f(R)n_f(z)$ yields the 
fermion density distribution
\begin{equation}
n_f(\tilde z;R)=\frac{2N_f}{\pi\lambda a_{zf}}
\left(1-\frac{\tilde z^2}{\lambda^2}\right)^{1/2},
\end{equation}
where $\lambda\simeq N_f^{1/2}[8R^3/(3\tilde g_{bf} N_b)]^{1/4}$ is 
the turning point for the fermions at large
interaction strength
$\tilde g_{bf}=|g_{bf}|/(a_{zb}\hbar\omega_{zf})$.
Finally, assuming that $\lambda> R$,
{\it i.e.} as long as the fermion cloud is much broader 
than the bosonic one, the total 
energy $E(R)$ becomes
\begin{eqnarray}
\frac{E(R)}{\hbar\omega_{zb}}&\simeq& 
\left(\frac{3m_f\omega_f^3}{2m_b\omega_b^3}\right)^{1/2}
(\tilde g_{bf}N_b)^{1/2}\frac{N_f^2}{4 R^{3/2}}
+\frac{3g_{bb}N_b^2}{10 R\,a_{zb}\,\hbar\omega_b}\nonumber\\
&+& 
\left(
\frac{3\tilde g_{bf}}{8}\right)^{1/4}
\frac{5 N_b^{5/4}N_f^{1/2}g_{bf}}{4\pi R^{3/4}a_{zf}\hbar\omega_b}+
\frac{\omega_fN_f^2}{\sqrt{24}\omega_b
(\tilde g_{bf}N_b)^{1/2}}R^{3/2}+\frac{N_b}{10}R^2.
\end{eqnarray}
The validity of this expression is limited by the diluteness condition
$2Ra_{zb}\gg N_b\,{\rm max}(a_{bb},|a_{bf}|)$.

\begin{figure}
\centerline{
{\epsfig{file=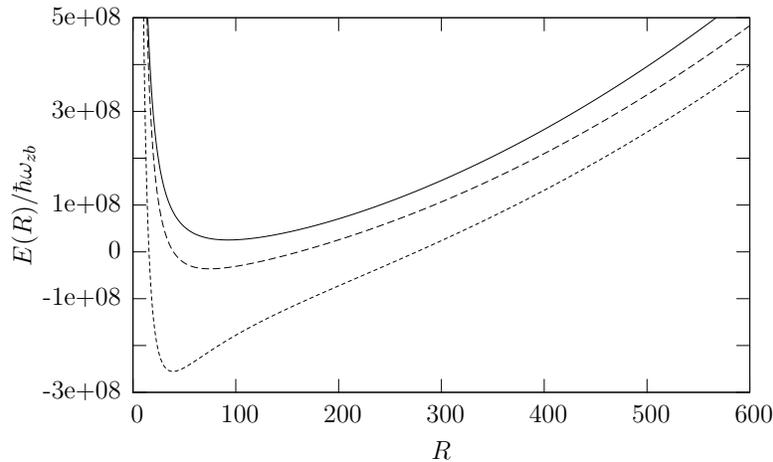,width=0.7\linewidth}}}
\caption{Energy $E(R)$ (in units of $\hbar\omega_{zb}$) of
a boson-fermion cloud in Q1D harmonic confinement,
as a function of the variational parameter $R$ for 
$N_b=N_f=10^4$.
The curves correspond to
$a_{bf}/a_\perp=-0.26$ (solid line), $-0.5$ (dashed line),
and $-1$ (short-dashed line).}
\label{fig_miy}
\end{figure}

\begin{figure}
\centerline{
{\epsfig{file=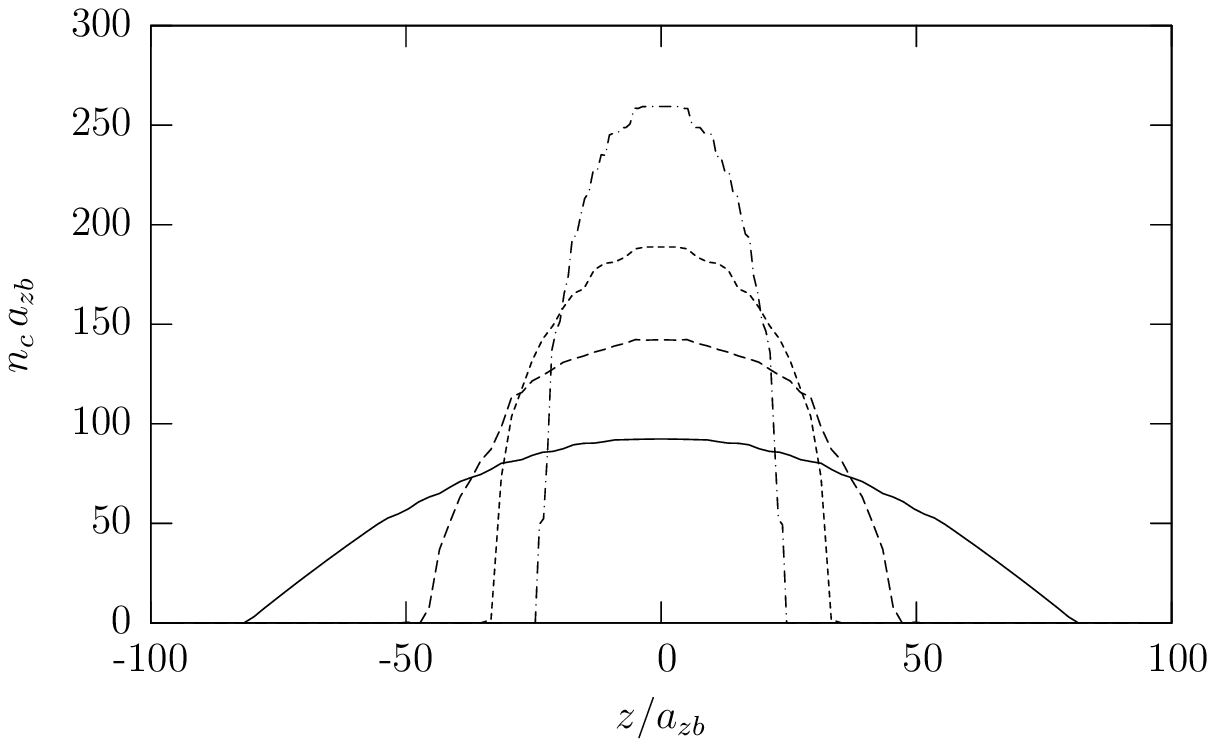,width=0.48\linewidth}}
{\epsfig{file=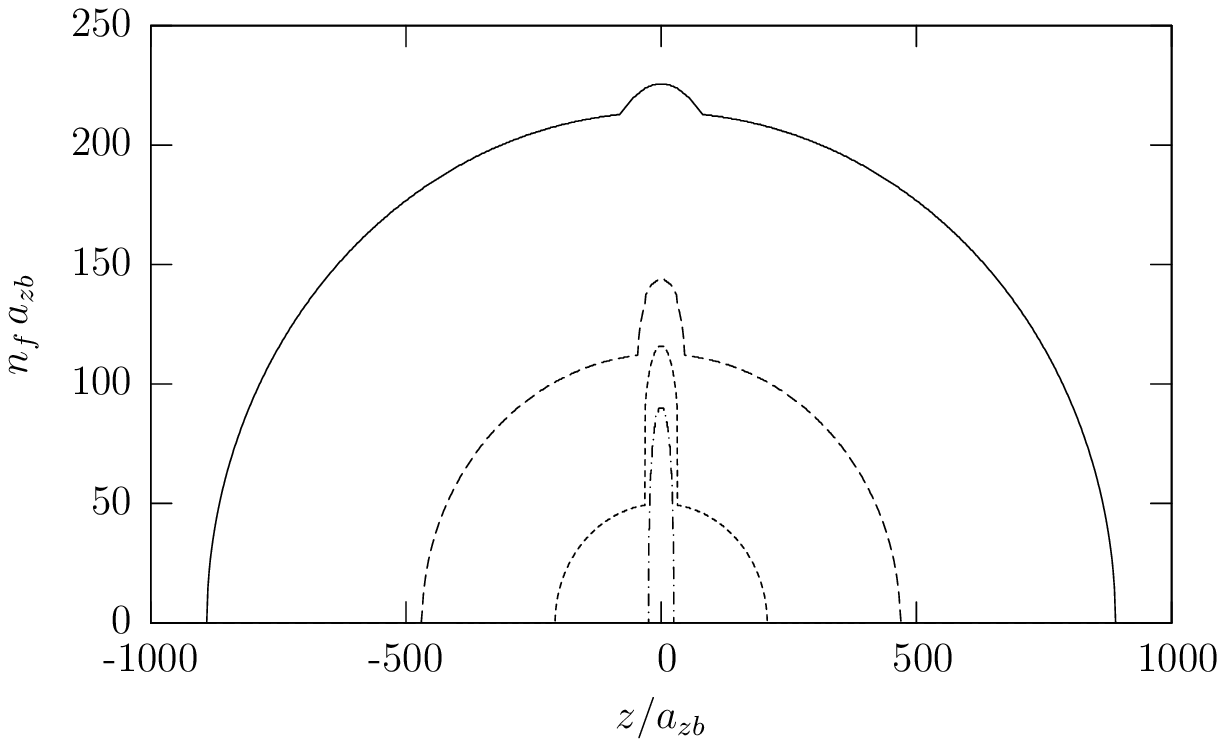,width=0.48\linewidth}}}
\caption{Density profiles of the condensate (left panel) and 
of the fermions (right panel) at
$a_{bf}/a_\perp=-0.5$ for various numbers of fermion atoms: 
$N_f=3\times 10^5$ (solid line), $N_f=8.5\times 10^4$ (long-dashed line), 
$N_f=2\times 10^4$ (short-dashed line), and $N_f=3740$ (dotted-dashed line).} 
\label{fig_coll}
\end{figure}

Contrary to the 3D case, in Q1D the boson-fermion attraction energy does 
not modify the leading term of $E(R)$ in the small-$R$ region,
so that $E(R)$ remains positive and for $R\rightarrow 0$
(see Fig. \ref{fig_miy}).
The mean-field 
model only leads to 
narrower and narrower density profiles, until pressure is relieved
by the expulsion of atoms from the trap through
three-body processes allowing collapse to proceed further.
Some illustrations of mean-field density profiles in correspondence
to the symbols shown in Fig. \ref{fig3} is given in Fig. \ref{fig_coll}.
The condensate density increases at the centre of the trap with
a decrease in
the number of fermions, which is thus seen to favour collapse.

\section{Summary and concluding remarks}
\label{secconcl}
In summary, we have evaluated a mean-field description of a 
boson-fermion mixture confined 
inside a cigar-shaped trap, such that the scattering events can still 
be considered as three-dimensional but nevertheless affected by the
radial confinement. 
We have focussed on the equilibrium properties of the mixture
and in the macroscopic limit
we have given a universal
phase diagram for the transitions to demixing or collapse 
expressed in terms of scaling parameters of the system.

We have studied the 
boson-fermion interaction energy in a mesoscopic cloud
as a function of
a boson-fermion repulsive coupling and
have given approximate analytical 
expressions identifying three critical regimes:
(i) partial demixing where the boson-fermion interaction energy 
attains maximum value from a balance between increasing interactions 
and diminishing overlap; 
(ii) dynamical demixing where the fermionic density drops 
to zero at the centre of the trap and a sharp dynamical signature of 
demixing may be expected; and 
(iii) full demixing where the boson-fermion overlap is negligible
as in the macroscopic limit. 
It is remarkable that in this model full demixing can be reached 
by simply {\it decreasing} the number of
trapped fermions. 
In the region of phase separation we have found 
several metastable configurations for the demixed cloud, having various 
topologies but lying at higher energy above the stable configuration 
which is composed of a core of condensed bosons surrounded by fermions. 

When the boson-fermion interaction is instead attractive, 
a Q1D mixture reaches collapse in the macroscopic limit
under the same conditions as it attains full demixing for
repulsive coupling, {\it i.e.} upon {\it lowering}
the fermion density. 
While a mean-field treatment is clearly unable to describe the full
path to collapse, we have seen that it supports the above view.
It would be interesting to study the Q1D 
mixture with strong attractive interactions by using models beyond mean-field 
and with inclusion of three-body collisions.

\vspace{1cm}
This work was partially supported by an Advanced Reasearch Initiative
of Scuola Normale Superiore (SNS).
One of us (Z.A.) thanks SNS for a Visiting Grant and 
acknowledges support from TUBITAK and from the Research Fund 
of the University of Istanbul under Project Number BYP-118/12122002.

%\bibliographystyle{jpb.bst}
%\bibliography{pr}

\end{document}